\documentstyle[12pt,aaspp4,graphics,epsf]{article}
\tighten
\slugcomment{To appear in ApJ Letters (accepted Jan 19, 1999)}
\def\solmass{$\hbox{M}_\odot$}
\def\solum {$\hbox{L}_\odot$} 
\def\kms {\hbox{${\rm km\, s}^{-1}$}} 
\def\percc {$\hbox{{\rm cm}}^{-3}$}    
\def\cmsq  {$\hbox{{\rm cm}}^{-2}$}    
\def\mic {$\mu\hbox{m}$}
\def\HII {\hbox{H{\sc ii}}}
\def\CEIO {\hbox{${\rm C}^{18}{\rm O}$}}   
\def\WAT {\hbox{${\rm H}_2{\rm O}$}}   
\def\SOTW  {\hbox{${\rm SO}_2$}}            
\def\FORM {\hbox{${\rm H}_2{\rm CO}$}}   
\def\MEFORM {\hbox{${\rm HCOOCH}_3$}}    
\def\CYAC {\hbox{${\rm HC}_3{\rm N}$}}     
\def\METH {\hbox{${\rm CH}_3{\rm OH}$}}  
\def\ETHCN {\hbox{${\rm CH}_3{\rm CH}_2{\rm CN}$}}  
\begin{document}
%
%
   \title{Hot gas and dust in a protostellar cluster near W3(OH)}

   \author{Friedrich Wyrowski   \altaffilmark{1,2}, 
           Peter Schilke        \altaffilmark{1}, 
           C. Malcolm Walmsley  \altaffilmark{3},
       and Karl M. Menten       \altaffilmark{1} }

\altaffiltext{1}{Max-Planck-Institut f{\"u}r Radioastronomie,
                 Auf dem H{\"u}gel 69, D-53121 Bonn, Germany }
\altaffiltext{2}{Department of Astronomy, University of Maryland,
                 College Park, MD 20742-2421}
\altaffiltext{3}{Osservatorio Astrofisico di Arcetri, 
                 Largo E. Fermi 5, I-50125 Firenze, Italy}
%
%
   \begin{abstract} 
     We used the IRAM Interferometer to obtain sub-arcsecond
     resolution observations of the high-mass star-forming region
     W3(OH) and its surroundings at a frequency of 220~GHz.  With the
     improved angular resolution, we distinguish 3 peaks in the
     thermal dust continuum emission originating from the hot core
     region $\approx 6$\arcsec\ (0.06 pc) east of W3(OH).  The dust
     emission peaks are coincident with known radio continuum sources,
     one of which is of non-thermal nature. The latter source is also
     at the center of expansion of a powerful bipolar outflow observed
     in \WAT\ maser emission. We determine the hot core mass to be
     15~\solmass\ based on the integrated dust continuum emission.
     Simultaneously many molecular lines are detected allowing the
     analysis of the temperature structure and the distribution of
     complex organic molecules in the hot core. From HNCO lines,
     spanning a wide range of excitation, two 200~K temperature peaks
     are found coincident with dust continuum emission peaks
     suggesting embedded heating sources within them.
   \end{abstract}

%

\newpage

 \section{Introduction}
 The archetypical ultracompact \HII\ region (UC \HII) W3(OH) has been
 the topic of numerous studies targeted at understanding the phenomena
 involved in the formation of massive stars and the interaction of
 these objects with their environment.  In particular, interferometric
 observations at radio and millimeter wavelengths have revealed an
 ever more detailed picture of the dense gas associated with W3(OH)
 (e.g., Baudry et al.\ 1993; Bloemhof et al.\ 1992; Reid et al.\ 1995;
 Turner \& Welch 1984; Wilner et al.\ 1995; Wink et al.\ 1994;
 Wyrowski et al.\ 1997).
 
 The observations suggest that there are several sites of very recent
 and on-going star formation: the most prominent of these are the UC
 \HII\ itself, which is ionized by a young O star, and a region
 showing hot molecular line and dust emission that is associated with
 strong water maser emission. The latter region is located at a
 projected distance of $\approx 0.06$~pc to the east of the UC \HII,
 assuming a distance of 2.2 kpc for W3(OH).  Most of the dense neutral
 gas in the W3(OH) complex appears to be associated with the \WAT\ 
 maser/``hot core'' source, whereas the most luminous young star in
 the region seems to be the exciting star of the UC \HII.
 
 There are several lines of evidence suggesting that there are one or
 more young stars embedded in the dense molecular gas near the \WAT\ 
 maser source: Turner \&\ Welch (1984) found a compact source of
 emission in the HCN $J = 1-0$ transition toward the \WAT\ maser
 position (in the following referred to as W3(OH)-TW).  The HCN line
 showed broad wings suggestive of mass loss from a young embedded
 star.  Alcolea et al.\ (1992) measured the \WAT\ maser proper motions
 and found them to be consistent with a bipolar outflow along an E-W
 axis in the plane of the sky.  Mauersberger et al.\ (1986a,b) used
 their ammonia observations to demonstrate the presence of dense, hot
 ($> 160$~K) gas close to the water masers and this was confirmed by
 the studies of Wink et al.\ (1994) and Helmich et al.\ (1994).
 However, the most surprising result was perhaps that of Reid et al.\ 
 (1995), who studied weak elongated centimeter continuum emission
 toward the TW source. They concluded that there was strong evidence
 for this being synchrotron emission from a ``jet-like'' structure
 aligned with the outflow indicated by the proper motions of the \WAT\ 
 masers associated with W3OH-TW.  Moreover, the synchrotron emission
 centroid was found to be coincident with the center of expansion of
 the \WAT\ outflow as determined by Alcolea et al.\ (1992).  All of
 these facts suggest the presence of a relatively luminous embedded
 star which drives a collimated outflow.
 
 In this paper, we present sensitive, sub-arcsecond resolution
 interferometric observations of the W3(OH) complex at a frequency of
 220 GHz; results from simultaneous observations at 107 GHz will be
 presented elsewhere. We find an excellent coincidence between the
 structures seen at millimeter wavelengths in thermal dust emission
 and those seen in non-thermal emission (see Wilner et al.\ 1998) at
 centimeter wavelengths.

 \section{Observations}
 
 Our observations were made with the 5 element Plateau de Bure
 Interferometer in three configurations: B1-N13 on 1998 January 6/7,
 A1 on February 7, and B2 on February 17. The phase center was
 $\alpha(J2000)$ = $02^{\rm h}27^{\rm m}04{\rlap.}{^{\rm s}}284$,
 $\delta(J2000)$ = +$61^{\circ}52'24{\rlap.}{''}55$, which is between
 the positions of W3(OH) and W3(OH)-TW. The total observing time on
 source was 17.1 hours covering a baseline range from 30 to 410~m. We
 used the dual-frequency receiver systems to simultaneously observe
 the \METH\ $J_k = 3_1-4_0A^+$ and \CEIO\ ($J = 2-1$) lines.  Due to
 good winter weather conditions, the 220 GHz system temperature was in
 the range 200 to 400~K and the radio seeing on all days better than
 0\farcs5. On-source integrations of 20~min were interspersed with
 phase calibrator observations on the quasar 0224+671.  For bandpass
 and flux density calibration, the sources 3C454.3, 3C273 and 3C84
 were used and the absolute flux density scale was established by
 observing MWC~349, for which a flux density of 1.66~Jy was assumed at
 220 GHz.  From the day-to-day variance in W3(OH)'s continuum flux
 density, we estimate that our 220 GHz flux density scale is accurate
 to within 20\%.
 
 The data were processed using the GILDAS software package.  To remove
 the effects of short-term atmospheric fluctuations on the
 interferometer phases, phase corrections derived from 220 GHz total
 power measurements were applied to the 107 and 220 GHz data (see
 Bremer et al.\ 1996).  Then, correcting for phase drifts on long
 timescales was done in the standard manner by observing a calibrator
 source. The phase solutions for the 107~GHz data were subtracted,
 appropriately scaled, from the 220~GHz data.  The RMS noise of the
 fits to the residual 220 GHz phase was found to be only 10\arcdeg\ on
 average.  After gridding and Fast Fourier Transform of the $uv-$data,
 a 220 GHz beam size of 0\farcs83$\times$0\farcs55 (FWHM) was
 determined for natural weighting with a position angle of
 100$\arcdeg$. From all correlator units, spectral line data cubes
 were built and checked for frequency ranges free of line emission in
 order to produce a continuum map. All data were deconvolved using the
 CLEAN algorithm and the resulting dynamical range limited RMS noise
 in the 220 GHz continuum maps is 20~mJy.

 \section{Results}

  \subsection{220 GHz continuum measurements}
  
  In Fig.~\ref{cmap}, we present our 220 GHz continuum map of W3(OH)
  superimposed upon the 8~GHz VLA map of Wilner et al.\ (1998).  For a
  better comparison with the latter, our 220 GHz map was restored with
  a circular beam of 0\farcs5 diameter (FWHM) and is thus slightly
  superresolved. On both maps, the main features are the UC \HII\ to
  the west and the water maser source 6\arcsec \ to the east.  We
  checked the alignment of the maps by computing their correlation and
  found them to coincide within 0\farcs05. Toward the \HII\ region,
  the main emission mechanism even at 220 GHz is free-free emission
  from the ionized gas. In fact, in agreement with the measurements of
  Wyrowski et al.\ (1997), we observe an integrated flux density of
  $3.1 \pm 0.5$~Jy at 220~GHz toward the \HII\ region as compared to a
  value of $3.0 \pm 0.15$~Jy at 107~GHz.  Combining the 3~mm
  measurements with results from the literature allows us to
  extrapolate the flux density to 220~GHz assuming optically thin
  free--free emission and we find $3.0 \pm 0.3$~Jy.  This value is
  consistent with the actually measured flux density at 220~GHz and we
  place a conservative upper limit of 0.5 Jy on any contribution from
  emission from dust associated with the UC \HII\ region.
  
  Most strikingly, Fig.~\ref{cmap} reveals the small-scale spatial
  coincidence of dust emission features (shown in grey-scale) with
  radio continuum sources detected toward the TW-object in the 3.6~cm
  map of Wilner et al.\ (1998, shown as contours). As seen on the
  inset to Fig.~\ref{cmap}, the easternmost component A is coincident
  with the centroid of the elongated structure seen with the VLA.
  There is also a rough correspondence between components B and C of
  the 220 GHz map with features seen at 3.6~cm.  In particular, it
  seems clear that dust emission source A is associated with the
  non-thermal radio source detected at centimeter wavelengths.
  
  We note here that there is also evidence in Fig.~\ref{cmap} for
  extended emission surrounding components A, B, and C.  Our
  observations are not sensitive to structures extended over size
  scales larger than about 5\arcsec\ and hence it is likely that we
  are missing flux.  We consider therefore that the total integrated
  flux density measured toward the water maser position (1.6~Jy from
  Fig.~\ref{cmap}) is a lower limit.  Estimates of the integrated
  fluxes in components A, B, and C are rendered difficult due to,
  both, the presence of the extended emission and blending obvious on
  Fig.~\ref{cmap}. Consequently, we only quote peak flux densities of
  components A, B, and C measured in our 0\farcs5 beam in
  Table~\ref{peaks}.

  \subsection{Line measurements near 220 GHz}
  
  Compared to previous, similar, observations, our data have much
  improved sensitivity and angular resolution and thus allow a new
  discussion of the molecular line emission distribution in the W3(OH)
  region.  The observations discussed in Wyrowski (1997) clearly
  showed that there is a dichotomy between oxygen-containing molecules
  related to methanol (essentially methyl formate and dimethyl ether)
  and nitrogen-containing species such as ethyl and vinyl cyanide.
  The former are seen both toward the UC \HII\ and the water masers
  whereas the latter are {\it only} seen toward the water masers.
  This is shown with improved angular resolution in
  Fig.~\ref{line-overlays} where we compare naturally weighted
  integrated intensity maps of lines detected near 220 GHz from a
  variety of species.  We note in particular the fact that the
  $24_{2,22}-23_{2,21}$ transition of \ETHCN \ and the $v_7=2$,
  $J=24-23$, transition of \CYAC\ (respectively 135~K and 772~K above
  ground) peak precisely in the direction of continuum component C and
  are not detected toward the \HII\ region. In contrast, the
  transitions which we detect from \METH, \MEFORM, and \FORM \ are
  seen both toward the ionized gas and toward the water maser
  concentration. The similarity of the intensity distributions in all
  three cases suggests a chemical link between these molecules as
  indeed discussed by Blake et al.\ (1987) in the context of Orion.
  Finally, we note the curious behavior of the $22_{2,20}-22_{1,21}$
  transition of \SOTW \ which, in contrast to e.g. \METH, is seen
  toward the eastern rather than the western border of the \HII\ 
  region.
  
  We also detected the $K_{a}=0,2,3,4$ components of the $J=10-9$ HNCO
  transitions at excitations of 50--750~K above ground.  The relative
  populations of these levels is thought to be determined by a
  combination of collisions and radiative transitions induced by the
  FIR radiation field within the hot cores (Churchwell et al.\ 1986).
  We have attempted to use the relative intensities of these lines as
  a ``thermometer'' measuring the temperature of the hot core by
  assuming, first, that all transitions are optically thin (which is
  justified by the fact that the $K_{a}=0$ line has a brightness
  temperature of less than 20~K) and, second, that the level
  populations are maintained by the radiation temperature of the dust.
  Since the dust is optically thick at the HNCO FIR pump wavelengths
  of 50--330~\mic, the radiation temperature equals the dust
  temperature.  We thus estimate rotational temperatures as a function
  of position over the water maser source.  In practise, the derived
  level column densities are consistent with a single rotational
  temperature at each point and allow us (assuming LTE) to infer the
  dust (and gas) temperature.  The resulting temperature map is shown
  in Fig.~\ref{hnco-results}, where one sees that there are two 200~K
  peaks roughly coincident with our continuum peaks A and C.  Between
  these there is a ``plateau'' where the temperature is of order
  150~K. We interpret this as evidence that there are two energy
  sources (young proto-B stars) embedded in the molecular gas whose
  positions are roughly defined by the temperature peaks and which are
  responsible for the cm-emission.
  
  Moreover, we can use the temperature distribution thus derived to
  deduce the dust and, thus, the hydrogen column density distribution
  from our continuum map.  The resulting hydrogen column density map
  is also shown on Fig.~\ref{hnco-results} where the resulting gas
  distribution appears to be {\it single-peaked} and extended with
  dimensions of $0.02\times 0.01$~pc. Using the formula given by
  Mezger et al.\ (1990) we calculate a total mass of 15~\solmass\ for
  this structure. Thus, the triple-peaked appearance of our continuum
  map (Fig.~\ref{cmap}) is caused by an interplay of the temperature
  and column density distributions.  This interpretation is supported
  by our \CEIO\ $(2-1)$ integrated intensity distribution
  (Fig.~\ref{hnco-results}) which has the same general form as our
  hydrogen column density map but is more sensitive to cooler material
  further from the embedded stars.

 \section{Discussion and Conclusions}
 
 Our new 220 GHz interferometer data suggest the presence of two
 embedded young stars lying slightly offset from the center of the gas
 clump associated with the water maser complex close to W3(OH). The
 mass of high temperature gas in the region is of order 15~\solmass,
 although our \CEIO \ map makes it clear that there is more cooler gas
 at larger distances from the protostars.  The luminosity of the
 hypothesised embedded stars coincident with positions A and C can be
 crudely inferred from the observed temperature distribution
 (Fig.~\ref{hnco-results}) by applying the Stefan-Boltzmann law. This
 way we find $3\times 10^4$~\solum\ for the gas clump with a factor of
 3 uncertainty due to temperature errors.  This (Panagia 1973)
 suggests embedded B0 stars.  It is of course likely that there are
 other lower mass objects associated with the complex and it is
 possible that there is more mass in stars than in gas.
 
 Finally, we note that perhaps the most remarkable result of this
 study has been the coincidence of continuum source A with the
 ``synchrotron jet'' of Wilner et al.\ (1998). What is the
 significance of this?  It seems reasonable that the dense hot core
 gas plays a role in confining the ``jet''. The magnetic pressure of
 the medium in which the relativistic electrons radiate ( B$^{2}/(8\pi) 
 \sim \, 4\, 10^{-6}$ erg \percc, Reid et al.\ 1995) is of the same
 order as the thermal pressure in the molecular gas for a density of
 $10^8$ \percc \ and thus such confinement seems feasible.  The origin
 of the relativistic electrons is unclear but there are analogous
 cases known. An example is the the jet in Cepheus A (Garay et al.\ 
 1996), whose radio emission is thought to be produced in shocks
 resulting from the interaction of the jet with the confining medium.
 The explanation in the case of W3(OH) may be similar.

 \acknowledgments
 
 We thank D. Wilner and M. Reid for providing the VLA data in advance
 of publication and an anonymous referee for helpful comments.  CMW
 acknowledges travel support from CNR grant 97.00018.CT02 and ASI
 grant ARS-96-66. He would also like to thank the Max Planck Institut
 f{\"u}r Radioastronomie for hospitality during the course of this
 work.


\bibliography{}
\bibliographystyle{astron}

\newpage

\begin{table}
\centering
\caption[peaks]
        {1.3~mm continuum peak positions toward W3(OH)-TW}
\vspace*{2mm}
\label{peaks}
\begin{tabular}{lcccc}
\hline
 Position & $\alpha$(J2000) & $\delta$(J2000) & $S_{220 {\rm GHz}}$ &
 $T_{\rm rot}$(HNCO) \\
          &                 &                 & (mJy~beam$^{-1}$) &  (K) \\
\hline
 A   & $02^{\rm h}27^{\rm m}04{\rlap.}{^{\rm s}}71$  &   
       $61^{\circ}52'24{\rlap.}{''}6$  & 185 & $200\pm 40$ \\
 B   & $02^{\rm h}27^{\rm m}04{\rlap.}{^{\rm s}}60$  &   
       $61^{\circ}52'24{\rlap.}{''}8$  & 145 & $150\pm 15$ \\
 C   & $02^{\rm h}27^{\rm m}04{\rlap.}{^{\rm s}}52$  &   
       $61^{\circ}52'24{\rlap.}{''}7$  & 150 & $190\pm 15$ \\
\hline 
\end{tabular}
\end{table}


\newpage

\figcaption[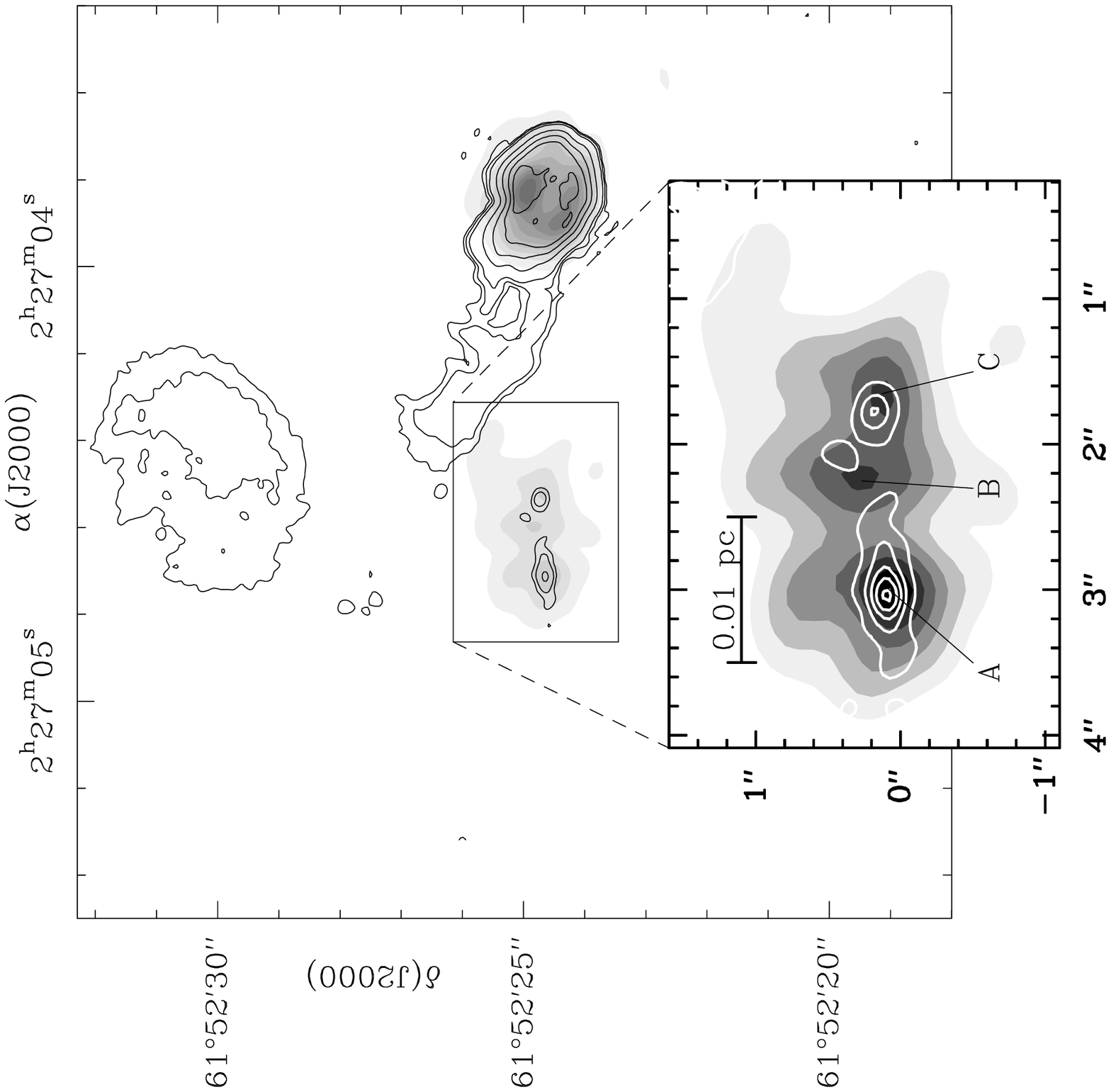]
  { 
    The large panel shows the 220~GHz Plateau de Bure continuum map of
    the W3(OH) complex in greyscale (47~mJy beam$^{-1}$ steps)
    superimposed on the 8.4~GHz VLA map (contours) of Wilner et al.\ 
    (1998). 220 GHz emission is observed from the UC \HII\ region
    W3(OH) to the west and the \WAT/TW region to the east.  The
    restoring beam is 0\farcs5 as compared to 0\farcs2 for the VLA
    measurements.  The inset depicts a blow-up of the \WAT/TW region
    showing the clumps A, B, and C referred to in the text (greyscale;
    steps of 24~mJy beam$^{-1}$, starting with 47~mJy beam$^{-1}$)
    with the 8.4 GHz contours overlaid.
  \label{cmap}
  }

\figcaption[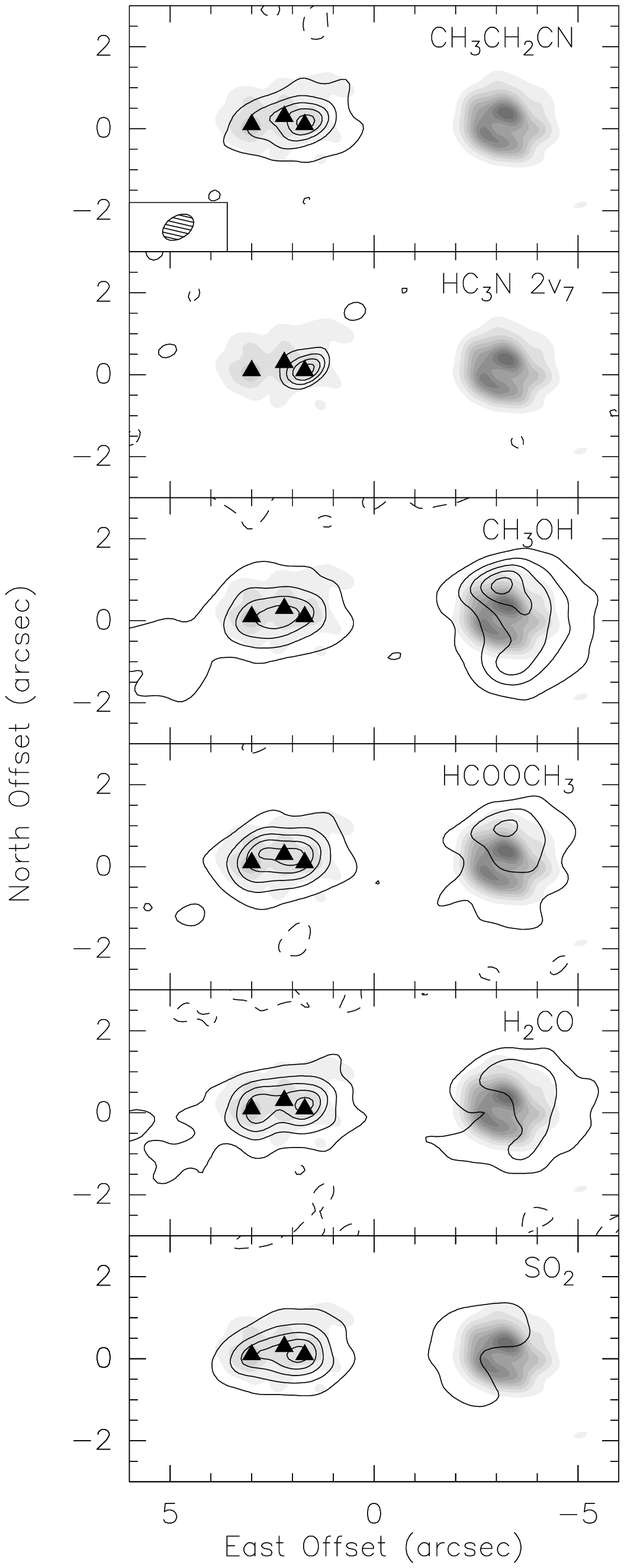]
  {
    Integrated intensity maps (contours, velocity windows of
    10--20~\kms) of the principal species detected in the 220 GHz
    band: (from the top) \ETHCN\ $24_{2,22}-23_{2,21}$, \CYAC\ $\,
    v_{7}=2$, $24-23$, \METH\ $5_{1,3}-4_{2,3}$, \MEFORM\ 
    $18_{4,15}-17_{4,14}$, \FORM\ $9_{1,8}-9_{1,9}$, \SOTW\ 
    $22_{2,20}-22_{1,21}$ superimposed on the 220 GHz continuum (see
    Fig.~\ref{cmap}) in grey scale.  Contour units are steps of 20\%
    of the peak emission.  The continuum peaks A, B, C
    (Fig.~\ref{cmap}) are marked by triangles. Position offsets are
    relative to the phase center position given in \S2.
  \label{line-overlays}
  }

\figcaption[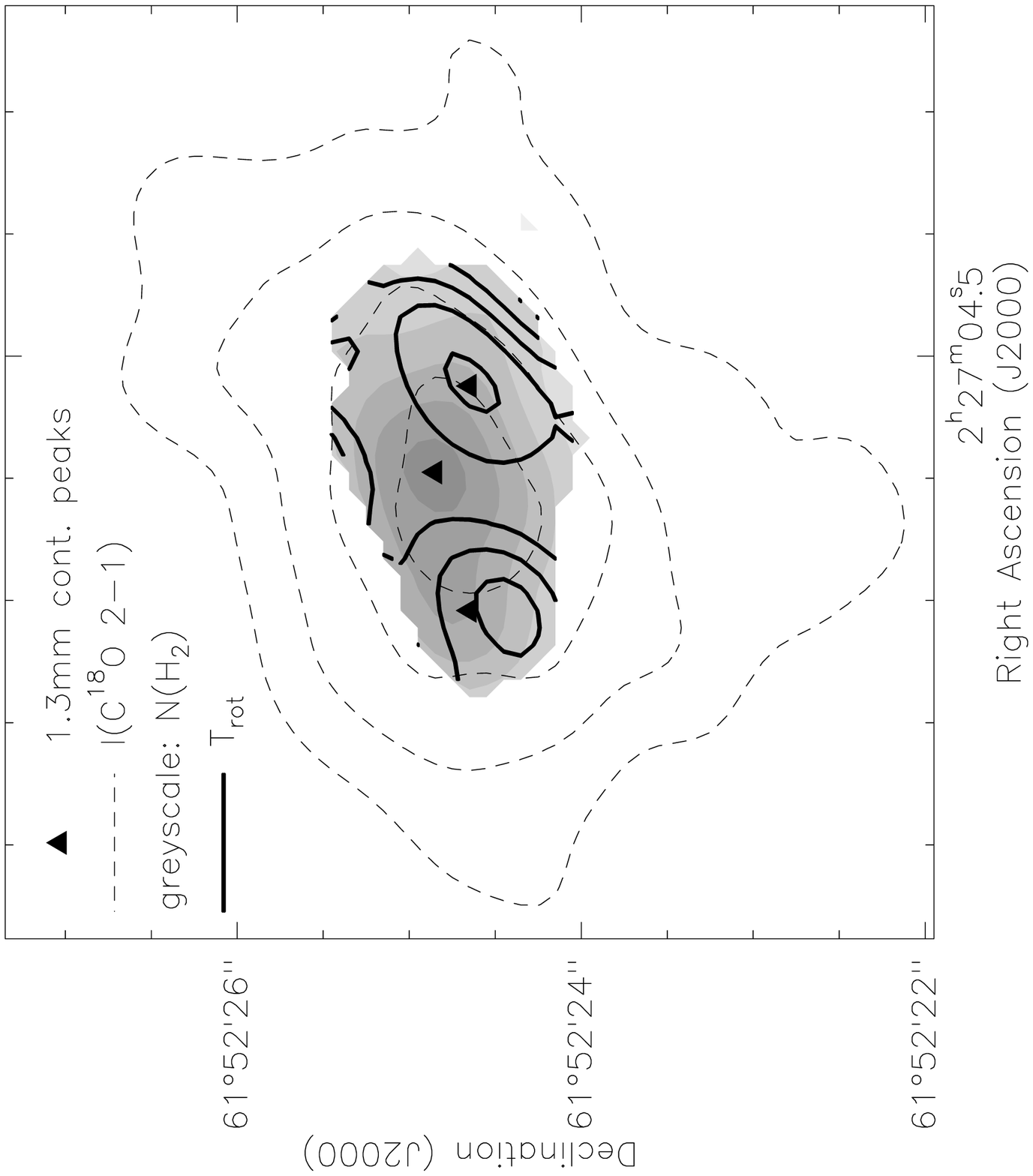]
  {
    Map of the rotation temperatures (thick contours: 160, 180, 200~K)
    derived from the HNCO $K_a=0, 2, 3$ lines. The molecular hydrogen
    column density map obtained by combining this temperature
    distribution with our natural weighting continuum map is shown in
    grey scale, from 1 to $3.5\,10^{24}$~\cmsq\ by
    $0.5\,10^{24}$~\cmsq. The dashed contours show for comparison with
    the latter our \CEIO(2--1) integrated intensity map with the same
    spatial resolution.  Contours are 54, 91, 127, 163~K\kms.
  \label{hnco-results}
  }

\newpage

\begin{figure}
  \epsfysize=\textwidth
  \rotatebox{-90}{\epsfbox{wyrowski-fig1.eps}}
\end{figure}

\begin{figure} 
  \hbox to \textwidth {
  \hfill  
  \epsfxsize=8cm
  \epsfbox{wyrowski-fig2.eps}
  \hfill}
\end{figure}

\begin{figure}
  \epsfysize=\textwidth
  \rotatebox{-90}{\epsfbox{wyrowski-fig3.eps}}
\end{figure}

\end{document}